# First-principles investigation on diffusion mechanism of alloying elements in dilute Zr alloys

Hai-Jin Lu [ab], Henry Wu [b], Nan Zou [a], Xiao-Gang Lu [ac], Yan-Lin He [a], Dane Morgan [b*]

[a] School of Materials Science and Engineering, Shanghai University, Shanghai 200072, China

[b] Department of Materials Science and Engineering, University of Wisconsin-Madison, Madison, Wisconsin 53706, United States

[c] State Key Laboratory of Advanced Special Steel, Shanghai University, Shanghai 200072, China

* Corresponding author. Email address: ddmorgan@wisc.edu

## Abstract

Impurity diffusion in Zr is potentially important for many applications of Zr alloys, and in particular for their use of nuclear reactor cladding. However, significant uncertainty presently exists about which elements are vacancy vs. interstitial diffusers, which can inhibit understanding and prediction of their behavior under different temperature, irradiation, and alloying conditions. Therefore, first-principles calculations based on density functional theory (DFT) have been employed to predict the temperature-dependent dilute impurity diffusion coefficients for 14 substitutional alloying elements in hexagonal closed packed (HCP) Zr. Vacancy-mediated diffusion was modeled with the eight-frequency model. Interstitial contributions to diffusion are estimated from interstitial formation and select migration energies. Formation energies for each impurity in nine high-symmetry interstitial sites were determined, including significant effects of thermal expansion. The dominant diffusion mechanism of each solute in HCP Zr was identified in terms of the calculated vacancy-mediated activation energy, lower and upper bounds of interstitial activation energy, and the formation entropy, suggesting a rough relation with the metallic radii of solutes. It is predicted that Cr, Cu, V, Zn, Mo, W, Au, Ag, Al, Nb, Ta and Ti all diffuse predominantly by an interstitial mechanism, while Hf, Zr, and Sn are likely to be predominantly vacancy-mediated diffusers at low temperature and interstitial diffusers at high temperature, although the identification of mechanisms for these elements at high-temperature is quite uncertain.

## 1. Introduction

Zirconium alloys have been widely used as cladding and core structural materials in the nuclear industry due to their combination of excellent neutron economy, good mechanical properties and significant corrosion resistance [1]. Impurity diffusion is a fundamental and important factor for understanding of many phenomena in Zr alloys, such as precipitation, creep and irradiation damage [2-4]. Essential to the modeling and control of impurity diffusion is to understand the mechanism by which impurity diffusion occurs. It is frequently the case that for pure metals with close packed structures like FCC and HCP, impurity diffusion occurs by a mono-vacancy mechanism. Based on this fact, first-principles methods have been employed to predict the impurity diffusivities in many close-packed hosts, such as Mg, Al, Co, Cu, Ni, Pd and Pt [5-7], generally yielding results which are in good agreement with experimental data when available. Nevertheless, the phenomenon of unexpectedly rapid impurity diffusion, generally denoted as fast and ultra-fast impurity diffusion, has been observed experimentally in HCP Zr for some solutes, including Fe, Co, and Ni [8-11], with impurity diffusivities being up to nine orders of magnitude faster than the self-diffusivity. Abnormal fast diffusion also occurs in many "open" metals [12] and has been extensively investigated in the literature. The interstitial diffusion mechanism is widely believed to be responsible for most of the observed fast diffusion. Frank and Turnbull first proposed a significant role for interstitial diffusion in what they called the dissociative mechanism [13], which proposed that solute atoms can be dissolved in not only substitutional but also to some extent interstitial sites, and that the measured diffusivity is therefore composed of contributions from both the vacancy-mediated and interstitial diffusion. In many cases, even a small fraction of interstitial dissolution can produce the observed fast diffusion.

Regarding the diffusion mechanism of impurities in HCP Zr, Tendler and Abriata [14] proposed the size-effect criterion for solutes to favor interstitial diffusion when the ratio of metallic radius for 12-fold coordination [15] of the impurity and solvent atoms in their pure states ($r_c$) is less than 0.85. However, this cutoff value is chosen empirically to separate slower and faster diffusing atoms, and there is no proof that it rigorously separates interstitial and vacancy diffusers. In particular, the solutes near the $r_c$ boundary, such as V, Zn, Mo etc. were regarded as vacancy diffusers since their diffusivities are relatively slow in comparison to that for, e.g., Fe and Co. However, this does not assure that these species near boundaries are actually vacancy diffusers as they could be some form of slow interstitial diffusers. Neumann and Tuijn also proposed size as the dominant factor in Pb and HCP Ti, and suggested that size impacts the diffusivity for interstitials mainly through its effects on impurity solubility. In their model, the common observation that smaller solute diffuse faster is assumed to be specifically due to an increased fraction of solute dissolved in interstitial sites [16]. Perez et al. summarized the solute diffusion in HCP Ti and Zr in terms of the impurities, the diffusion anisotropy, the atomic size effect and the correlation between solubility and diffusivity. As with other authors, Perez et al. suggest that impurity diffusion in HCP Ti and HCP Zr is characterized by fast interstitial diffusion of small atoms and slow substitutional diffusion of large atoms [17]. Beside the uncertainty in the impurity diffusion mechanism in Zr, Hood [12] also pointed out that experimental diffusion data displaying abnormal behaviour in HCP Zr can be ascribed to extrinsic effects, such as grain boundaries, dislocations and impurity-enhanced diffusion. The latter is particularly a concern since Zr materials are notoriously difficult to purify, as they aggressively dissolve Fe and oxygen. The first-principles calculations performed by Pérez and Weissmann [18] verified that the presence of impurity Fe creates a considerable lattice distortion together with an increase in the number of vacancies, and thus leads to an enhancement in the self and substitutional diffusion in HCP Zr. Given these uncertainties in intrinsic diffusion mechanisms and the possible role of extrinsic factors in the present data, the diffusion mechanism for impurities in HCP Zr, especially those with sizes close to $r_c$, is still an open question that would benefit from additional investigation.

Some insight into what we expect for the behavior of impurities in Zr can be gained by considering recent studies quite similar to the present work of impurities in Ti, another hcp metal in the same column of the periodic table as Zr. Specifically, Bernstein et al. calculated the formation energies to analyze the normal and fast diffusion of some solutes in HCP Ti [19] and Zhang et al. identified the dominant diffusion mechanism for some solutes in HCP Ti [20] based on the calculated defect formation energies and migration energies. Both these studies suggest that smaller solutes are more likely to diffuse by interstitial mechanism, consistent with the Zr case. In particular, they together suggest that Fe, Co, Ni, Cu, Mn and V are interstitial diffusers and Si, Ga, Al, Au, Nb, In, Hf, Zr, Sn and Pb are likely vacancy diffusers, although for some of these vacancy diffusers the mechanism is quite uncertain. The atomic size ratios [15] of these elements relative to Ti are 0.85-0.91 and 0.94-1.16 for the set of interstitial and vacancy-mediated diffusers, respectively, suggesting a crossover from interstitial to vacancy dominated diffusion mechanisms for a size ratio range of 0.91-0.94. Compared to Zr, this value is somewhat higher than that proposed by Tendler and Abriata [14] of 0.85, but similar to the value of 0.92-0.99 we find in this work (see details below).

In the present work, the vacancy-mediated impurity diffusion coefficients of 14 representative solutes near and off the limit of radius ratio of 0.85 (Cr, Cu, V, Zn, Mo, W, Al, Au, Ag, Nb, Ta, Ti, Hf, Sn) in HCP Zr were calculated using the eight-frequency model [21]. We employed the climbing image nudged elastic band method (CI-NEB) [22-23] to find the minimum energy path of atomic hops. The Vineyard's harmonic transition state theory [24] was utilized to calculate the attempt frequencies for substitutional migrations. To identify the dominant diffusion mechanism of each solute, the formation energies of each solute residing in nine high-symmetry interstitial sites of HCP Zr were also calculated as well as select interstitial migration barriers for the subsequent estimation of the activation energies for interstitial diffusion. By comparing the vacancy-mediated and interstitial activation energies, the dominant diffusion mechanism of each solute in HCP Zr was identified.

## 2. Diffusion models

The eight-frequency model developed by Ghate [21] was utilized to calculate the vacancy-mediated impurity diffusivity in the present work. As shown in Fig. 1, eight frequencies are specified to evaluate the diffusion coefficient in the model. The $\omega_a$ and $\omega'_a$ are the vacancy-solute rotation jumps between two adjacent basal planes and vise versa. The $\omega_b$ and $\omega'_b$ are vacancy-solute rotation jumps within basal and $c$ axis planes. The $\omega_c$ and $\omega'_c$ are vacancy-solute dissociation jumps within basal and $c$ axis planes. The $\omega_X$ and $\omega'_X$ are the vacancy-solute exchange jumps within basal and $c$ axis planes. The two unique solute jumps within basal plane and between two adjacent basal planes, resulting from anisotropy in the HCP lattice, lead to two distinctive diffusion coefficients, $D_\perp$ and $D_\parallel$, which denote the diffusivities perpendicular and parallel to the $c$ axis, respectively. The interactions of solutes with other solutes are assumed to be negligible due to the dilute solute concentrations, and they are neglected in this model.

### 2.1. Vacancy concentration

The vacancy formation energy without solute atoms $E_F^V$ is determined by the energy difference of the perfect structure and that with monovacancy:

$$E_F^V = E_{Zr_{N-1}Va_1} - \frac{N-1}{N} E_{Zr_N} \qquad (1)$$

where $N$ denotes total atoms in the structure, $Va$ represents a vacancy.

The solute-vacancy binding energy plays an important role in determining the activation energy, which is defined as follow [25-26]:

$$E_B = E_{Zr_N} + E_{Zr_{N-2}X_1Va_1} - E_{Zr_{N-1}X_1} - E_{Zr_{N-1}Va_1} \qquad (2)$$

where $X$ indicates a substitutional solute atom in the Zr supercell. Note that with this definition a negative binding energy represents a favorable binding between solute and vacancy. The formation energy for a substitutional solute atom in HCP Zr is determined by:

$$E_F^S = E_{Zr_{N-1}X_1} - \frac{N-1}{N}E_{Zr_N} - E_X \quad (3)$$

where the $E_X$ means the total energy of a solute atom in its ground state crystal structure at room temperature.

The concentration of vacancy with an adjacent solute atom, $C_{V\text{-}S}$, can be expressed by the vacancy formation energy, vacancy-solute binding energy and the corresponding vacancy formation and binding entropy $S_F^V$ and $S_B^V$, through the following Arrhenius equation:

$$C_{V\text{-}S} = \exp\left(\frac{S_F^V + S_B^V}{k_B}\right)\exp\left(-\frac{E_F^V + E_B}{k_B T}\right) \quad (4)$$

where $k_B$ is the Boltzmann constant and $T$ means temperature. The vacancy formation entropy (vibrational) was adopted from previous work [27] as 3.19 $k_B$. The binding entropy $S_B^V$ is assumed to be negligible and not calculated in the present work.

## 2.2. Jump frequency

Each jump frequency in the eight-frequency model can be evaluated by the following formula:

$$\omega_i = v_i \exp\left(-\frac{E_i}{k_B T}\right) \quad (5)$$

where $E_i$ is the migration barrier and $v_i$ expresses the attempt frequency for the jump. The CI-NEB method with a single intermediate image was employed to calculate the substitutional migration barriers for atomic jumps. One image is assumed to be sufficient because the migrations occur between nearest-neighbor close-packed lattice sites, and therefore are expected to have only a single maximum. In the present work, we approximated the full prefactor determination by using only two prefactors, one for all solute atom transitions and the other for all solvent atom transitions. These prefactors were determined for the solute and solvent by calculations of migration of the solute for an isolated vacancy-solute pair and migration of a solvent atom in pure solvent, respectively. The attempt frequency $v_{hop}$ was calculated based on Vineyard's harmonic transition state theory [24], which formally requires calculations of phonon frequencies for all atoms. However, we approximated Vineyard's full expression by using only the phonon frequency of the hopping atom in its initial position, $v^{ini}$, and at its saddle point configuration $v^{saddle}$:

$$v_{hop} = \frac{\prod_1^{3n} v_i^{ini}}{\prod_1^{3n-1} v_i^{saddle}} \sim \frac{\prod_1^{3} v_i^{ini}}{\prod_1^{2} v_i^{saddle}} \qquad (6)$$

The approximations of using just two prefactors in applications of Eq. 5, and determining those factors from just the hopping atom vibrations (see Eq. 6), are expected together to lead to errors of less than 10x in overall vacancy-mediated diffusion coefficients, primarily because they capture essential physics and the relatively small overall variability expected in the prefactors. Similar approximations were shown to be quite accurate in previous work [5]. Therefore, these approximations should not impact any of the qualitative conclusions of this paper, but could have a significant impact in the quantitative predictions.

**2.3. Eight-frequency model**

When a solute atom exchanges its position with an adjacent vacancy, it has a significant possibility to jump back to its origin position subsequently. Consequently, the atom does not diffuse as expected for a completely random series of jumps. Generally, the motions of atoms are not completely independent of previous jumps except for the interstitial diffusion mechanism. The correlation factor $f$ was introduced to measure the non-randomness of the atomic motion process, which depends on both the diffusion mechanism and lattice geometry.

To evaluate the expressions for the correlation factors of solute diffusion in an HCP structure, Ghate [21] introduced the vector $S_i$ as the average final displacement of the tracer after $i$th jumps. Owing to the anisotropy in an HCP lattice, there are two different kinds of jump length, $\lambda_A$ for non-basal jumps and $\lambda_B$ for basal jumps. The $\lambda_{Ab}$ corresponds to the projection of $\lambda_A$ on the basal plane. The final displacement along $x$ and $y$ axis in a same basal plane denoted as $S_{Bx}$ and $S_{By}$ and parallel to $c$ axis as $S_{Az}$ as well as its projection on the basal plane $S_{Ab}$, are obtained from the following equations:

$$S_{Bx} = \frac{\sqrt{3}\omega_a S_{Ab} + \omega_b S_{Bx} - \omega_X(\lambda_B + S_{Bx})}{2\omega_a + 2\omega_b + 5.512\omega_c + \omega_X} \tag{7}$$

$$S_{By} = \frac{\omega_a S_{Ab} - \omega_b S_{By}}{2\omega_a + 2\omega_b + 5.512\omega_c + \omega_X} \tag{8}$$

$$S_{Az} = \frac{-\omega'_X(S_{Az} + 1)}{2\omega'_a + 5.512\omega'_c + \omega'_X} \tag{9}$$

$$S_{Ab} = \frac{\omega'_a(\sqrt{3}S_{Bx} + S_{By}) - \omega'_b S_{Ab} - \omega'_X(\lambda_{Ab} + S_{Ab})}{2\omega'_a + 2\omega'_b + 5.512\omega'_c + \omega'_X} \tag{10}$$

The correlation factors for jumps used in Ghate's eight-frequency model depend on the jump frequencies and displacements as follows:

$$f_{Az} = \frac{2\omega'_a + 5.512\omega'_c}{2\omega'_a + 5.512\omega'_c + 2\omega'_X} \tag{11}$$

$$f_{Bx} = 1 + \frac{2S_{Bx}}{\lambda_B} \tag{12}$$

$$f_{Ab} = 1 + \frac{2S_{Ab}}{\lambda_{Ab}} \tag{13}$$

where the subscripts represent the directions as used for displacements. Then, the two different diffusion coefficients $D_\perp$ and $D_\parallel$ can be evaluated based on the following equations [28]:

$$D_\parallel = \frac{3}{4} C_{V\text{-}S} c^2 f_{Az} \omega'_X \tag{14}$$

$$D_\perp = \frac{1}{2} C_{V\text{-}S} a^2 (3 f_{Bx} \omega_X + f_{Ab} \omega'_X) \tag{15}$$

where $a$ and $c$ express the lattice parameters of an HCP structure. $C_{V\text{-}S}$ represents the concentration of vacancy adjacent to the solute, as defined previously.

We note that recently a more accurate multi-frequency model for hcp impurity substitutional diffusion has been developed by Agarwal and Trinkle [29-30]. This new model was published after we had completed much of the present work and so is not used in this paper. Comparison between Ghate's model used here for Mg impurity diffusion and the new model show almost identical activation energy values for most metals and even when values differed they were within 0.1 eV. Therefore, we think that the present model is adequate for the conclusions in this work and that a more accurate multi-frequency approach would have no qualitative, and likely only a minor quantitative effect.

**2.4 Interstitial configurations**

To investigate the activation energy for interstitial diffusion, we performed the calculations for each solute atom residing in nine high-symmetry interstitial configurations displayed in Fig. 2, namely tetrahedral (T), hexahedral (H), octahedral, (O), basal octahedral (BO) sites, crowdions in and out of the basal plane (BC and C), split dumbbells along the *c* axis and diagonal direction (S and DS) and in the basal plane (BS). By considering this wide-range of sites, and by including small symmetry breaking perturbations of the initial positions in select cases, we believe that we have captured all the stable interstitial sites in the lattice for each solute. The formation energy of a solute atom in an interstitial site of HCP Zr is calculated by the expression:

$$E_F^I = E_{Zr_N X_1} - E_{Zr_N} - E_X \qquad (16)$$

where *X* stands for a solute atom and *N* means the total number of atoms in the perfect structure. Interstitial migration barriers between two adjacent O within the basal plane and along *c* axis direction were determined by CI-NEB method with three images. Note that O may not be the most stable interstitial configuration for some solutes.

**3. Computational details**

In the present work, the Vienna Ab initio Simulation Package (VASP) code [31-32] was employed to perform the electronic structure calculations within the framework of the density functional theory. The projector-augmented wave (PAW) method was used to model the electron-ion interaction [33]. The Generalized Gradient Approximation (GGA), as parameterized by Perdew, Burke, and Ernzerhof (PBE) [34] was adopted to describe electron exchange and correlation. All calculations were performed with the spin-polarized approach. The plane wave cutoff energy of 350 eV was chosen for the plane-wave expansion of the electronic wave functions. For the HCP structure, a 4×4×3 supercell containing 96 atoms was chosen for all the bulk and defect calculations. The undefected perfect cell was fully relaxed and then only cell-internal (no lattice parameter or volume) ionic relaxations were performed for all the additional studies. An energy convergence of $10^{-5}$ ($10^{-2}$) eV/atom was used for all electronic (ionic) relaxation calculations. A Gamma centered mesh of 5×5×5 was used for the electronic integration in the Brillouin zone. The convergence tests for these values can be found in previous work [5]. The high-throughput workflow software MAterials Simulation Toolkit (MAST) [35] developed at University of Wisconsin-Madison based on the pymatgen toolset [36] was utilized to automate the workflows of all the calculations.

## 4. Results and discussions

The self-diffusion coefficient of HCP Zr has been widely measured experimentally by many researchers [37-42]. However, as can be seen in Fig. 3, these experimental results vary significantly due to the uncertainties introduced by extrinsic effects, likely dominated by the presence of impurities in the sample materials, especially Fe and oxygen. The pre-exponential factor and activation energy determined by Hood and Schultz based on the melting point rule [39] are $5\times10^{-5}$ m$^2$/s and 2.85 eV, respectively. These values show good agreement with some of the diffusion values from Horvath et al. [37] over a wide temperature range, supporting their accuracy. However, the data from Horvath et al. [37] show a strong downward curvature with increasing $1/T$, which was proposed to be due to impurity enhancement, as discussed in Ref. [43]. Data from Hood et al. [41] with very low impurities gives a lower diffusion coefficient and higher activation energy of 3.17 eV for $D_{||}$, but this is based on just three data points over a relatively small temperature range of 163 K. The calculated temperature-dependent self-diffusivities $D_{\perp}$ and $D_{||}$ of HCP Zr from DFT in this work are presented in Fig. 3, which fall within the scattered experimental data, demonstrating a good agreement with the general experimental trends. Our calculated pre-exponential factors and activation energies are $1.65\times10^{-5}$ m$^2$/s and 1.87×10$^{-5}$ m$^2$/s, 2.58 eV and 2.65 eV for $D_\perp$ and $D_\parallel$. Note that the approach of correction used in Ref. [5] to improve the capability of predicting the vacancy-mediated diffusivity was not employed in this work, primarily due to the large uncertainties in Arrhenius fits to diffusion coefficients from experimental results and the ambiguity in the dominant diffusion mechanism of HCP Zr.

The vacancy concentration in a system is exponentially related to the formation energy, which has important consequences on the microscopic transportation of atoms and then plays an important role in determining the vacancy-mediated diffusivity, as in Eq. 14 and 15. The experimental measurements of the vacancy formation energy in HCP Zr are rather scarce. Hood [44] used the absence of positron trapping at vacancies to infer an upper limit for vacancy concentration and thereby a corresponding lower limit for the formation energy of 1.35 eV at 1136 K. The present calculated vacancy formation energy of HCP Zr is 2.01 eV, in good agreement with previous predictions tabulated in Table 1. The consistent calculated values suggest that the experimental inference of 1.35 eV by Hood is significantly below the true value, which we take to be our calculated value of 2.01 eV in the models developed in this work.

The interaction between a solute atom and an accompanying vacancy is characterized by the solute-vacancy binding energy, which modifies the concentration of the vacancy adjacent to a solute atom and thereby influences the rate of atomic position exchanges. Strong binding between solute and vacancy might also affect the precipitation behaviour due to vacancy trapping. The solute-vacancy binding energy is difficult to measure experimentally but can be readily calculated by first-principles method. In the present work, we predicted the binding energies, $E_B$, for 14 solutes in HCP Zr via DFT calculations, which are all reported in Table 2. The positive values imply repulsive interaction while negative values imply attractive interaction. As seen in the Table 2, the Cr, Mo and W show a strong attraction to a vacancy while Al and Zn are strongly repulsive to a vacancy. The interactions between other solutes and vacancy are all less than or equal to 0.07 eV in magnitude, which is close to $k_BT$ at temperatures where diffusion is likely to be relevant, and therefore unlikely to play a significant role in modifying the diffusion.

The migration barriers of eight atomic hops demonstrated in Fig. 1 for 14 solutes in HCP Zr were calculated via CI-NEB method with one image. An atypical behavior sometimes seen for oversized solute diffusion, which requires a new formulation of correlation effects [48] is not observed in the present work since the metallic radii of solutes considered are relatively small as compared to host Zr, with the size ratio for the largest solute Sn being 1.02. The results shown in Table 2 indicate that the direct solute-vacancy migration barriers ($E_X$ and $E'_X$) are higher than that of the solvent migration barriers for most solutes. $E'_b$ is the lowest barrier for all solutes except for Cr, Cu and Au, for which the lowest barrier is $E'_a$.

Based on the foregoing calculated energies as well as the attempt frequencies for solute and solvent atoms presented in Table 3, the vacancy-mediated diffusion coefficients for all 14 solutes in HCP Zr were computed by employing the formulae of the eight-frequency model described previously. The activation energies and pre-exponential factors of each solute element were obtained by fitting the diffusion coefficients in the reciprocal temperature range of 0 to 20. They were collected in Table 4 and compared with available experimental data [39, 41, 42, 49-68]. The calculated activation energy for the diffusion parallel to *c* axis is always greater than that perpendicular to *c* axis, indicating a consistent anisotropy for solute diffusion in HCP Zr. Some care must be taken in making the comparison to experiments so we discuss the different cases in some detail here. The experimental measurements of Sn diffusion in HCP Zr has abnormally small activation energy and pre-exponential factor as described in Ref. [69], and appear to be outside the range one would expect for solute diffusion with relatively large radius in HCP Zr based on the other values collected in Ref. [69]. We therefore believe the experimental data on Sn is not representative of bulk volume diffusion in a pure Zr crystal and do not attempt to compare our results to Sn. Among the other solutes considered here, we found a good agreement between calculated and experimental diffusivities for Ta and Hf as demonstrated in Fig. 4. For most other solutes, especially those with relatively small metallic radii, such as Cr, Cu and Zn, the calculated vacancy mediated diffusivities are always slower than the experimental results. The deviation for Cr and Cu is readily interpretable in terms of an interstitial mechanism given their ultra-fast diffusion in HCP Zr (about seven orders of magnitude faster than the self-diffusion) as these elements have quite small metallic radii. More generally, we note that the experiments are frequently subject to uncertainties associated with the inevitable presence of impurities Fe and oxygen in the Zr samples, which tend to increase the solute diffusivity significantly. Therefore, it is possible that the consistently higher predicted experimental diffusivities compared to calculations for many solutes results at least in part from impurity effects present in the experiments. However, we predicted significantly slower diffusion through a vacancy-mediated mechanism than is observed for a number of impurities and across many experiments, suggesting that the discrepancy is not simply an impurity effect. The results suggest that, in contradiction to Tendler and Abriata's rule [14], some of the impurities with size ratio > 0.85 may move by an interstitial mechanism. To give insight into the impurity diffusion mechanism in HCP Zr, we investigated the possibility of interstitial diffusion for all solutes by calculating the formation energies of solute atoms dissolved in interstitial sites and select interstitial migration energies for estimation of the interstitial diffusion activation energies.

The formation energies and stability of self-interstitial configurations of HCP Zr were investigated first. The results show the most stable configuration is BO with the formation energy of 3.0 eV, while T is not a stable structure and decays to S after relaxation. The BC decays to BO and C decays to C′, which is a low-symmetry configuration formed by displacing the central atom of C toward an O site, as suggested by Vérité et al. [70]. The BO, O, BS and C′ are the four most stable self-interstitial configurations of HCP Zr, all having formation energies within 0.2 eV of one another. Similar calculations were performed for the 14 solutes considered in this work. The most stable interstitial configuration is O for Ta, Cr, W, Mo and Nb; BO for Ti, V and Hf; BS for Ag; DS for Al, Zn and Cu; C′ for Au; and C for Sn. The reason for the largest interstitial site O not being the most stable for some solutes can be ascribed to the chemical interaction proposed in Ref. [20]. The formation energies of vacancy-mediated and interstitial solutes in HCP Zr were tabulated in Table 5 for comparison. Some values are not available and marked NA as the configurations are not stable. It can be seen that all solutes show a preference for the substitutional site, with lower formation energies relative to the interstitial sites. In the present work, the thermal expansion effect on stabilization of the interstitials was investigated by calculating the formation energy of the most stable interstitial configuration for each solute found for zero temperature at a lattice parameter expanded from our zero-temperature value so as to match the thermal expansion expected at 1073 K. Thermal expansion was taken as $5.5\times10^{-6}$ deg$^{-1}$ along the $a$ axis, $10.8\times10^{-6}$ deg$^{-1}$ along the $c$ axis from Ref. [71]. The formation energy difference of the most stable interstitial and substitutional configurations at 0 K and 1073 K was collected in Table 6, where it can be seen that it is easier to form an interstitial site at high temperature. The effect of the thermal expansion is to reduce all the values by an average of 0.40 eV, with a standard deviation in the reduction of just 0.05 eV. The similar trend was observed in the HCP Ti from DFT calculations [20]. However, the recent molecular dynamics study of HCP Zr [72] suggests similar interstitial formation energy of 3 eV under 2000 K.

Due to the potentially complex interstitial energy landscape for the solutes, there is no tractable approach to quantitatively predict interstitial diffusion coefficients of all the impurities studied in this work. However, it is possible to determine a lower and upper bound for the interstitial diffusion activation energetics (or equivalently, an upper and lower bound for interstitial diffusion coefficients), as well as an intermediate value we take as a best estimate for interstitial diffusion activation energies, and use these to determine dominant solute diffusion mechanisms. Specifically, we calculated the migration barriers for O→O pathways within basal plane and along $c$ axis direction via CI-NEB method with three images for all solutes, which are presented in Table 7. The intermediate structure for O→O pathways within basal plane and along $c$ axis direction are C and BO respectively. The O→BO→O and O→C→O sequences therefore constitute simple pathways that can lead to a long-range diffusion. Note that the activated states occur along the pathways, not necessarily at the BO or C sites, so the barriers are equal to or higher than the $E_{BO}$-$E_O$ and $E_C$-$E_O$ values. The formation energy of the most stable interstitial configuration is a lower bound for the interstitial activation energy, since this is the value that would be obtained if the most stable interstitial could migrate with essentially zero migration barrier. If it is assumed that the O→O migration energy is representative for other interstitial migration energies, then adding this value to the lowest interstitial formation energy can be regarded as a reasonable estimation of the interstitial activation energy. We call this value our "best estimate" of the interstitial migration energy. The upper bound of the interstitial activation energy was obtained by including any energy needed to move the interstitial to the O interstitial site, and then adding the O→O migration barriers. As this mechanism provides a possible diffusion mechanism it must provide an upper bound on the interstitial activation energy, since any additional mechanisms would either have higher activation energy, in which case the interstitial O→O diffusion would dominate, or have lower activation energy, in which case the O→O diffusion is an upper bound. We note that the "best estimate" value will be equal to the upper bound when interstitial O site is the most stable. As we will show below, the lower and upper bound are usually enough to identify the active diffusion mechanisms.

We assume that the combined vacancy-mediated and interstitial diffusion operate by the dissociative mechanism [13], which operates for solutes incorporated not only on substitutional sites but also in the interstitial sites of the host, and has been used to account for fast diffusion in HCP Zr and other "open" metals. Under this mechanism the overall effective impurity diffusivity, $D_{eff}$, will be a weighted average of the vacancy-mediated and interstitial mechanisms, which can be expressed by the following form:

$$D_{eff} = \frac{C_S}{C_S + C_I} D_S + \frac{C_I}{C_S + C_I} D_I$$
$$= \frac{C_S}{C_S + C_I} D_0^S \exp\left(-\frac{Q_S}{k_B T}\right) + \frac{C_I}{C_S + C_I} D_0^I \exp\left(-\frac{E_M^I}{k_B T}\right) \quad (17)$$

where $D_S$, $D_0^S$ and $D_I$, $D_0^I$ denote the diffusion coefficients and the pre-exponential factors of the vacancy-mediated and interstitial diffusion, respectively. The $Q_S$ is the vacancy-mediated activation energy. The $E_M^I$ represents the migration barrier of the interstitial atomic hops. $k_B$ is the Boltzmann constant. The $C_S$ and $C_I$ stand for the concentration of solutes dissolved on substitutional and in interstitial sites respectively, which are given as:

$$C_S = \exp\left(\frac{S_F^S}{k_B}\right)\exp\left(-\frac{E_F^S}{k_B T}\right), \quad C_I = \exp\left(\frac{S_F^I}{k_B}\right)\exp\left(-\frac{E_F^I}{k_B T}\right) \quad (18)$$

where $E_F^S$ and $E_F^I$ ($S_F^S$ and $S_F^I$) are the substitutional and interstitial formation energy (entropy), respectively. For the situation that the substitutional solubility dominates (i.e. $C_S \gg C_I$), the effective diffusivity in Eq. 17 approaches:

$$D_{eff} = D_0^S \exp\left(-\frac{Q_S}{k_B T}\right) + D_0^I \exp\left(-\frac{E_I + E_M^I - S_F T}{k_B T}\right) \quad (19)$$

where $E_I = E_F^I - E_F^S$ ($S_F = S_F^I - S_F^S$) is the interstitial and substitutional formation energy (entropy) difference, which is the energy (entropy) of forming an interstitial solute from a substitutional one. In the present work, the interstitial pre-exponential factor $D_0^I$ was not calculated since they are expected to be roughly at the same order of magnitude as $D_0^V$. If we assume we can set $D_0^I = D_0^V$, then the values of $D_{eff}$ in Eq. 19 are wholly governed by the relative values of the corresponding activation energies and the formation entropy difference in the exponentials. The formation entropy difference $S_F$ is computationally too demanding to obtain with ab initio methods for the many solutes treated here and therefore was not calculated in this work. However, $S_F^I$ is often larger than $S_F^S$ due to the larger number of degrees of freedom for interstitial relative to substitutional species. Specifically, $S_F$ is predicted to be about 7 $k_B$ in Ref. [72] for pure Zr (relative to an $S_F^S$ taken as zero), and comparable self-interstitial formation entropies $S_F^I$ have been predicted for other systems [73-75]. Thus this term will likely increase the role of the interstitial diffusion mechanism relative to that of substitutional diffusion.

Fig. 5 shows the difference between the vacancy-mediated activation energies obtained from the eight-frequency model and the estimated interstitial activation energies. The solutes are ordered along the horizontal axis in increasing order of the metallic radii [15] of solutes. The vacancy-mediated diffusion perpendicular to the $c$ axis with relatively lower activation energy was adopted for comparison. The comparison using the activation energy of diffusion parallel to $c$ axis can be found in the Supplementary data Fig. 1 and yields qualitatively the same conclusions. All data in Fig. 5 and Supplementary data Fig. 1 can be found in the Supplementary data Table 1.

First we consider the prediction of interstitial vs. vacancy-mediated diffusion mechanisms based on the activation energies. As can be seen in Fig. 5a and 5b, if we use the best estimate values for the interstitial activation energies, all the elements smaller than Hf (these are Cr, Cu, V, Zn, Mo, W, Al, Au, Ag, Nb, Ta and Ti, as the plot is in order of increasing size) are predicted to be interstitial diffusers. For all but Au and Ag this is a robust prediction since the upper bound of interstitial activation energies are lower than the vacancy-mediated counterparts at both temperatures. For Au and Ag, the comparison at 0 K shows about 0.3 eV lower activation energy for vacancy-mediated than the upper bound interstitial diffusion. However, by 1073 K the interstitial diffusion is preferable under all models, as shown in Fig. 5b. Thus our results strongly suggest Au and Ag will be interstitial diffusers, although some uncertainty does remain and more detailed calculations would be needed for a definitive prediction. The entropy formation difference $S_F$ that we do not explicitly model in this work (see Eq. 19) is expected to further increase the dominance of the interstitial mechanism for these elements, which will have no impact on the qualitative mechanism identified, except to further increase the likelihood that Au and Ag are interstitial diffusers. The next three larger elements, Hf, Zr and Sn, are predicted to be vacancy diffusers at low temperature, since the lower bound of interstitial activation energies are about 0.4 eV higher than their vacancy-mediated counterparts at 0 K. $S_F$ is expected to play a more minor role than enthalpy in the dominant mechanism at low temperature since, when compared to enthalpy, its contribution is of the form $S_F T$ (see Eq. 19). Therefore, $S_F$ is expected to have a minor effect on the identification for Hf, Zr and Sn at low temperature. For Hf and Zr at high temperature, the lower bound, best estimate, and upper bound barriers for interstitials are still higher than the vacancy-mediated barriers, although the lower bound are quite close. Both the upper bound and best estimate interstitial barriers for Sn are much higher than vacancy-mediated diffusion, but the lower bound is quite close to substitutional diffusion barrier, making the prediction of dominant mechanism at high temperature somewhat ambiguous. However, the entropy term could play a significant role in the dominant diffusion mechanism with increasing temperature. Specifically, if we take the value of 7 $k_B$ from Ref. [72] as a reasonable estimate for $S_F$, the interstitial contribution to overall diffusivity will increase by about three orders of magnitude, or equivalently, $S_F T$ = 0.65 eV at 1073 K. Under this assumption Hf, Zr, and Sn would all be dominated by the interstitial mechanism by about 1000 K due to the formation entropy difference. Therefore, more detailed calculations or measurements will be needed to determine a robust dominant mechanism for these elements at high temperatures.

We also note that there is a systematic error for defect formation energy due to the so-called "electronic surface error" associated with special areas of the electronic density [76-80] that might significantly affect the present classification. The corrections for this error found in Ref. [78-79] are about 0.15-0.5 eV for a given defect formation energy. Furthermore, previous calculations have also shown that the interstitial formation energies for Si and Al differ from benchmark diffusion quantum Monte Carlo (QMC) calculations by about 1 eV and 0.3 eV, coming out higher than those from GGA and closer to experimental results [76-77]. If we assume that the bulk of these errors are due to similar problems in the physics modeling the activation energies of both vacancies and interstitials, then it is reasonable to expect that the errors will have similar qualitative trends for both activation energies. Thus the relative activation energies for vacancy vs. interstitial mechanisms may benefit from significant cancellation and be smaller than either error on its own. Although we cannot rigorously bound this error, taking the range from Ref. [78-79] (up to 0.5 eV) and from QMC vs. GGA (up to 1 eV) [76-77] and assuming some cancellation suggests that 0.5 eV is a reasonable upper bound error estimate. Shifting relative activation energies of vacancies and interstitials by this value would lead to a modest but significant qualitative impact on the conclusion. Specifically, if we stabilize the interstitials by 0.5 eV then the only qualitative changes is that the dominance of interstitial mechanism for impurities Hf, Zr and Sn will be increased at high temperature, and the prediction of mechanism for them at low temperature will be quite ambiguous due to rather close activation energies for both mechanisms. If we destabilize the interstitials by 0.50 eV, the qualitative prediction for impurities Al, Au, Ag, Nb and Ta at low temperature may have a significant or even dominant vacancy-mediated diffusion contribution. At high temperature, the best estimate values will be slightly lower than the vacancy-mediated diffusion, however, they are more likely interstitial diffusers at high temperature due to the expected contribution of the interstitial formation entropy. Hf, Zr and Sn will still be controlled by vacancy-mediated mechanism without uncertainty at low temperature and minor uncertainty at high temperature approaching 1073 K. However, assuming these worst case errors do not occur, we believe that the predictions of dominant mechanisms in this work are robust, although the exact activation energies likely have significant errors. In summary, these results suggest the first 12 solutes in Fig. 5 to be interstitial diffusers with minor uncertainty for Al, Au, Ag, Nb and Ta associated with the exchange-correlation functional approximations and a dominance of vacancy-mediated diffusion for Hf, Zr and Sn at low temperature with a possible but not certain crossover to interstitial mechanism at higher temperatures approaching 1073 K.

Now that the mechanistic predictions are clear, we can compare the results to other models and previous experiments collected in Table 4. Recently, Mendelev and Bokstein [72] employed interatomic potentials and molecular dynamics simulations to suggest that the self-diffusion of HCP Zr is dominated by the interstitial mechanism. Given their molecular dynamics calculations were performed in the range 800-2000 K, this result is consistent with our prediction of the dominant mechanism at temperatures approaching 1073 K. The calculated diffusion coefficients using eight-frequency model for Hf and Zr show a good agreement with experimental data as shown in Fig. 4. The experimental data for Sn is abnormal as described previously and thus is not used for comparison. For the interstitial diffusers at 0 K, the best estimate values for Cu, Zn, Al and Ag show an excellent agreement with experiments within deviation of 0.1 eV and the deviation is about 0.3 eV for Cr, V and Mo. However, the best estimate values for Nb, Ta and Ti are about 0.6 eV lower than the experimental data. At 1073 K, the best estimate values for all solutes are lower than the experimental results. The deviation for V and Mo is within 0.2 eV, about 0.4 eV for Cu, Zn, Al and Ag and 0.61 eV for Cr. The results for Nb, Ta and Ti are about 0.9 eV lower than the experimental data. The discrepancy between the estimated and experimental results can have contributions from both the DFT approaches, typically dominated by errors from the exchange-correlation functional approximations, and the experimental errors. As one simple way to explore possible exchange-correlation functional errors from the DFT, we calculated the formation energies for the most stable interstitial configuration of Nb, Ta and Ti by employing the Local Density Approximation (LDA) in place of the GGA used otherwise in this work. The results are collected in Supplementary data Table 2 and show similar trends, with 0.2-0.3 eV smaller interstitial formation energies from the LDA compared with those from the GGA. This result suggests that a simple change from GGA to LDA does not improve the predictions, but instead would make the underestimation of the barrier worse. As discussed previously, this might be attributed to the systematic underestimation of interstitial energies from the DFT that impacts both LDA and GGA approaches. The deviation in the high-temperature predictions range from 0.05-0.99 eV, which is somewhat larger than the range of errors of 0.15-0.5 eV associated with the corrections used in Ref. [78-79], but similar to the up to 1 eV scale of errors seen when comparing results of QMC to GGA [76-77]. Besides the DFT error, experimental error may also contribute to some of the discrepancy. As seen in Table 4, most experimental activation energies were obtained by fitting the data to a narrow temperature range about 100-200 K, which can give rise to large uncertainties in the activation energies. Also, the results from different authors vary significantly, for example, the experimental activation energies for Nb from Ref. [42] and Ref. [62] are 1.37 and 2.69 eV, respectively.

According to the foregoing discussions, the dominant mechanism of impurity diffusion in HCP Zr is roughly associated with the size effect, in that solutes with small metallic radii tend to diffuse predominantly by an interstitial mechanism and vice versa. However, our results suggest that the sharp classification of the metallic radius ratio of 0.85 proposed by Tendler and Abriata [14] is not correct, and that V, Zn, Mo, W, Al, Au, Ag, Nb, Ta and Ti, which were classified as vacancy-mediated diffusers in HCP Zr, are likely to be interstitial diffusers, although with some uncertainties for Al, Au, Ag, Nb and Ta at low temperature due to possible errors from the exchange-correlation functionals in the DFT. Our predictions suggest that a transition from interstitial to vacancy-mediated mechanism occurs with increasing size in the size ratio range 0.92-0.99.

## 5. Conclusions

First-principles calculations were employed to calculate the vacancy-mediated impurity diffusion coefficients for 14 solutes in HCP Zr as well as the self-diffusivity based on the eight-frequency model. The formation energies of solutes in nine high-symmetry interstitial configurations of solvent atoms were calculated, indicating a preference for substitutional site over interstitial site for all solutes. The thermal expansion effect will increase the stability of interstitial configurations. The select interstitial migration barriers of O→O within basal plane and along *c* axis direction were calculated. The lower bound, best estimate and upper bound of interstitial activation energies were determined based on the calculations. Based on the comparison of the calculated vacancy-mediated and estimated interstitial activation energies, the dominant diffusion mechanism of each solute in HCP Zr was analyzed, which is roughly related to the size-effect criterion. We predicted Cr and Cu to be interstitial diffusers, which is consistent with previously proposed size-effect criterion from Tendler and Abriata. Hf, Zr and Sn were regarded as vacancy-mediated diffusers previously and were predicted to be vacancy-mediated diffusers at low temperature and more likely interstitial diffusers at high temperature. We also identified the diffusion of V, Zn, Mo, W, Al, Nb, Ta and Ti in HCP Zr occurs predominantly by an interstitial mechanism, which were classified as vacancy-mediated diffusers previously. The errors from entropy effects that were not included in the calculations and the exchange-correlation functionals used in the DFT could have a significant impact on quantitative values and could easily lead to predictions of Hf, Zr and Sn at high temperature and Al, Au, Ag, Nb and Ta at low temperature as either interstitial or vacancy-mediated diffusers. The transition from interstitial to vacancy-mediated mechanism is predicted to occur in the atomic size ratio range of 0.92-0.99 instead of the 0.85 proposed previously by Tendler and Abriata. These results suggest that interstitial diffusion mechanism in close-packed crystal structures may be much more common than is generally assumed, although more studies are needed. Direct comparison to experimental activation energies suggests a consistent underprediction in the range of 0.05-0.99 eV, perhaps due to the so-called "electronic surface error" arising from limitations of the exchange-correlation functionals in the DFT.

## Acknowledgments

Xiao-Gang Lu was supported by National Key R&D Program of China (Grant number: 2017YFB0701502). Support for Dane Morgan was provided by the NSF Software Infrastructure for Sustained Innovation (SI2) award No. 1148011. Hai-Jin Lu is grateful to the financial support from China Scholarship Council (CSC). Computational resources for this work came from the UW-Madison Center For High

Throughput Computing (CHTC) and Advanced Computing Initiative (ACI) in the Department of Computer Sciences.

Fig. 1 The schematic illustration of atomic jumps in an HCP lattice required for the eight-frequency model of Ghate [21].

Fig. 2 Schematic representation of nine high-symmetry interstitial configurations in an HCP structure investigated in the present work.

Fig. 3 The calculated and corrected self-diffusion coefficients of HCP Zr compared with experimental results [37-42].

Fig. 4 The DFT calculated diffusivities compared with experimental data [39, 41, 42, 49-68], the solid and dash lines denote the $D_\perp$ and $D_\parallel$ respectively. The elements are placed in the order of their metallic radii size from (a) to (d)

Fig. 5 The difference between the vacancy-mediated activation energies (perpendicular to *c* axis) and the lower bound, best estimate and upper bound of the interstitial activation energies at (a) 0 K and (b) 1073 K.

Fig. 6 The difference between the calculated and experimental activation energies at (a) 0 K and (b) 1073 K.

Table 1 The comparison of present and previous DFT calculated vacancy formation energy and lattice parameters of HCP Zr.

**Table 2** The calculated solute-vacancy binding energies and migration energies of eight atomic exchanges demonstrated in Fig. 1 for solutes in HCP Zr. The data in bold denote the lowest migration barrier for each solute. Unit in eV.

**Table 3** The calculated attempt frequencies (Thz) for solute and solvent atoms

**Table 4** The predicted vacancy-mediated activation energies and pre-exponential factors for 14 solutes in HCP Zr compared with available experimental data.

**Table 5** The formation energies (eV) for substitutional and interstitial solutes in HCP Zr at ground state. Some of the values are not available due to the instability of the configurations and are marked NA, the data in bold means formation energy for the most stable interstitial site at ground state.

**Table 6** The formation energy difference of the most stable interstitial and substitutional configurations at 0 K and 1073 K. Unit in eV.

**Table 7** The interstitial migration energies (eV) for O→O pathways within basal plane and along *c* axis direction

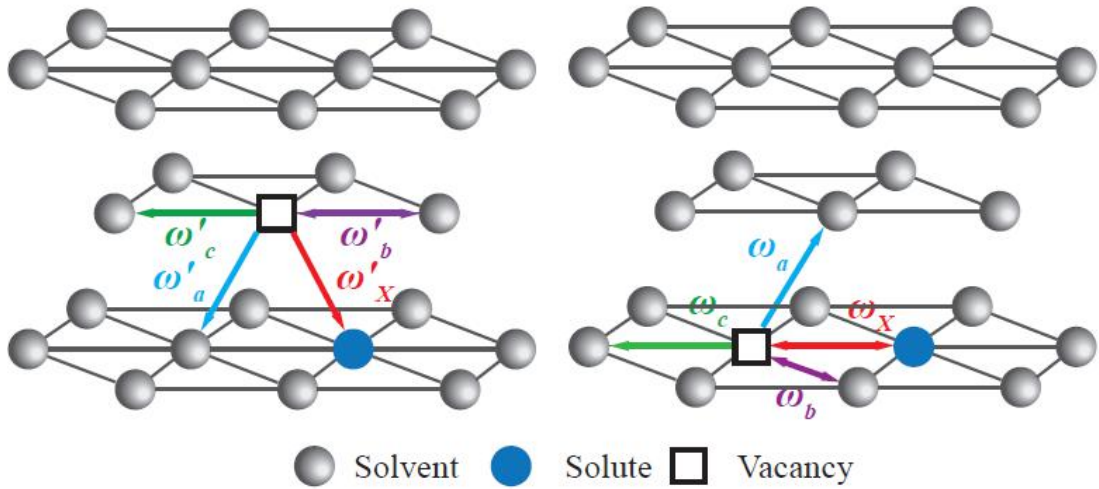

**Fig. 1** The schematic illustration of atomic jumps in an HCP lattice required for the eight-frequency model of Ghate [21].

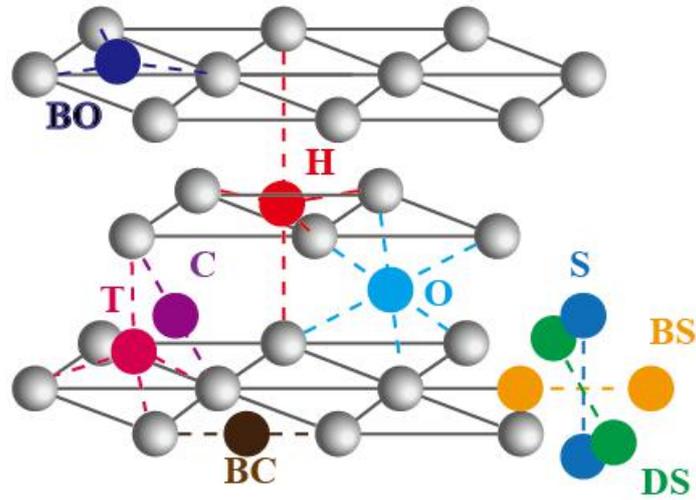

**Fig. 2** Schematic representation of nine high-symmetry interstitial configurations in an HCP structure investigated in the present work.

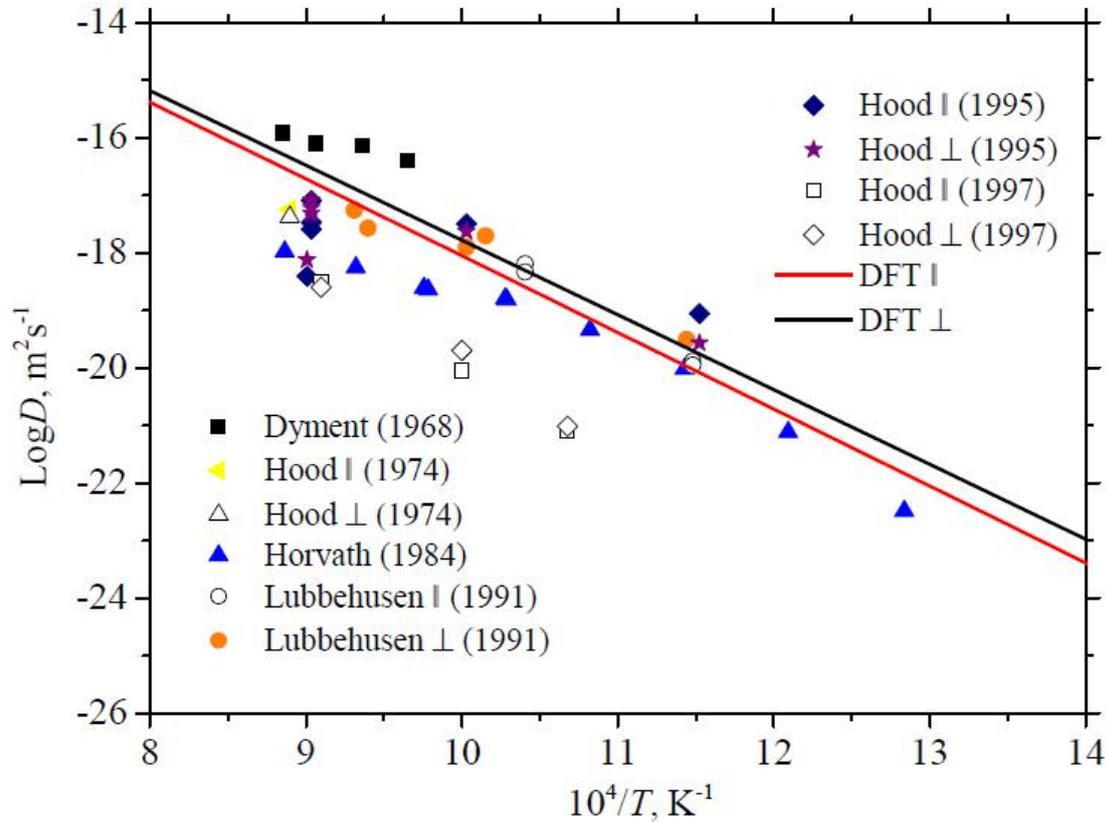

**Fig. 3** The calculated and corrected self-diffusion coefficients of HCP Zr compared with experimental results [37-42].

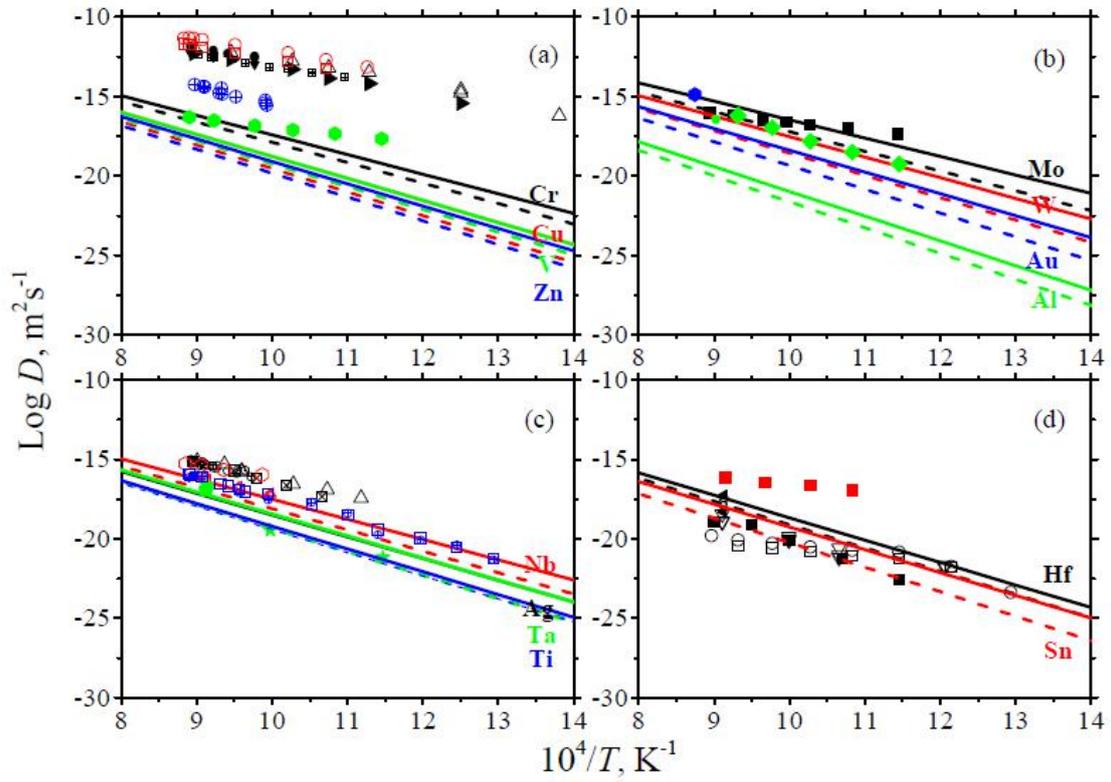

Fig. 4 The DFT calculated diffusivities compared with experimental data [39, 41, 42, 49-68], the solid and dash lines denote the $D_\perp$ and $D_\parallel$ respectively. The elements are placed in the order of their metallic radii size from (a) to (d)

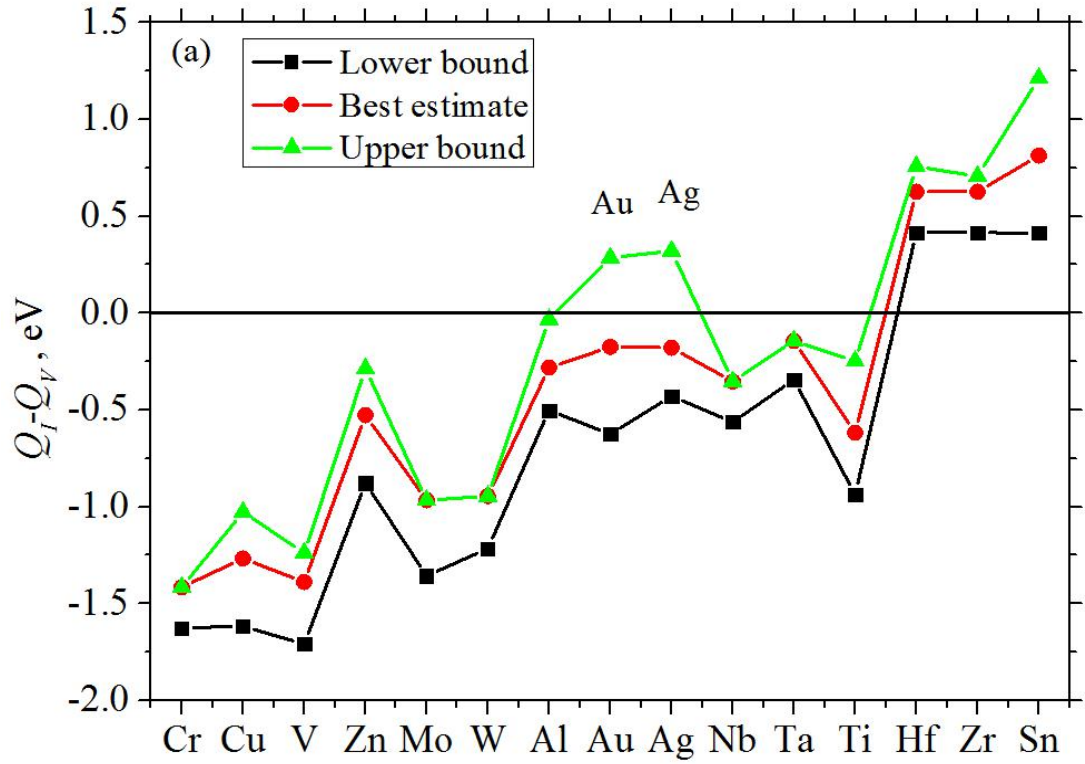
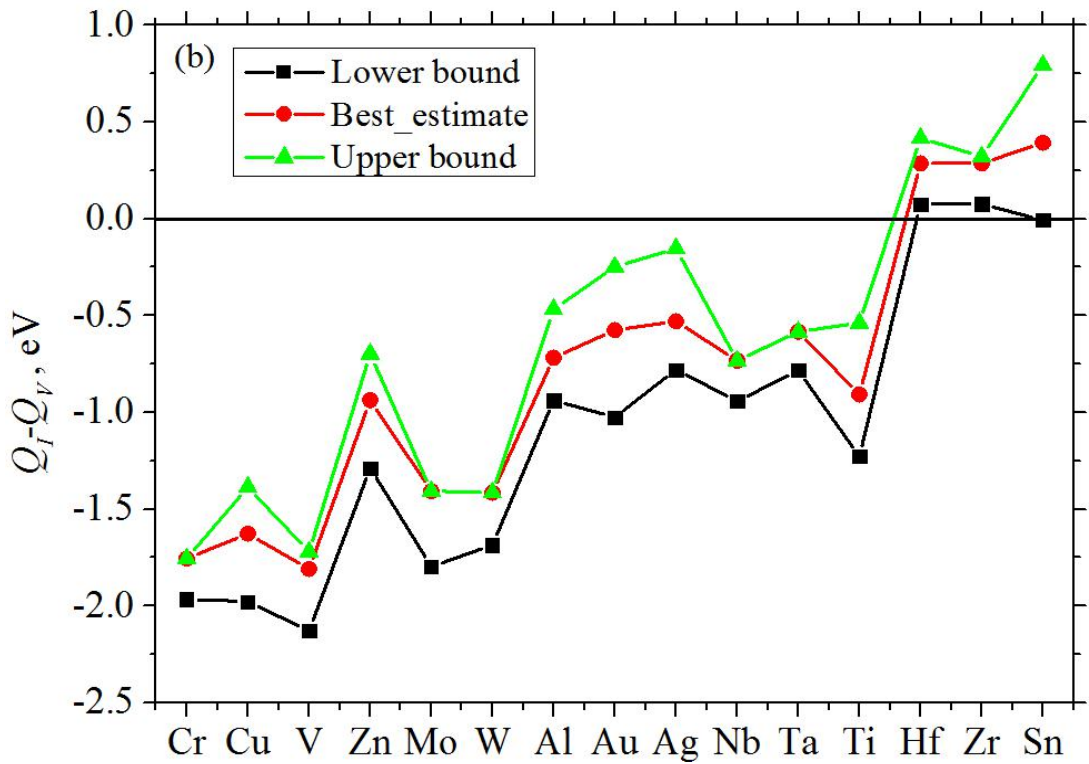

Fig. 5 The difference between the vacancy-mediated activation energies (perpendicular to *c* axis) and the lower bound, best estimate and upper bound of the interstitial activation energies at (a) 0 K and (b) 1073 K.

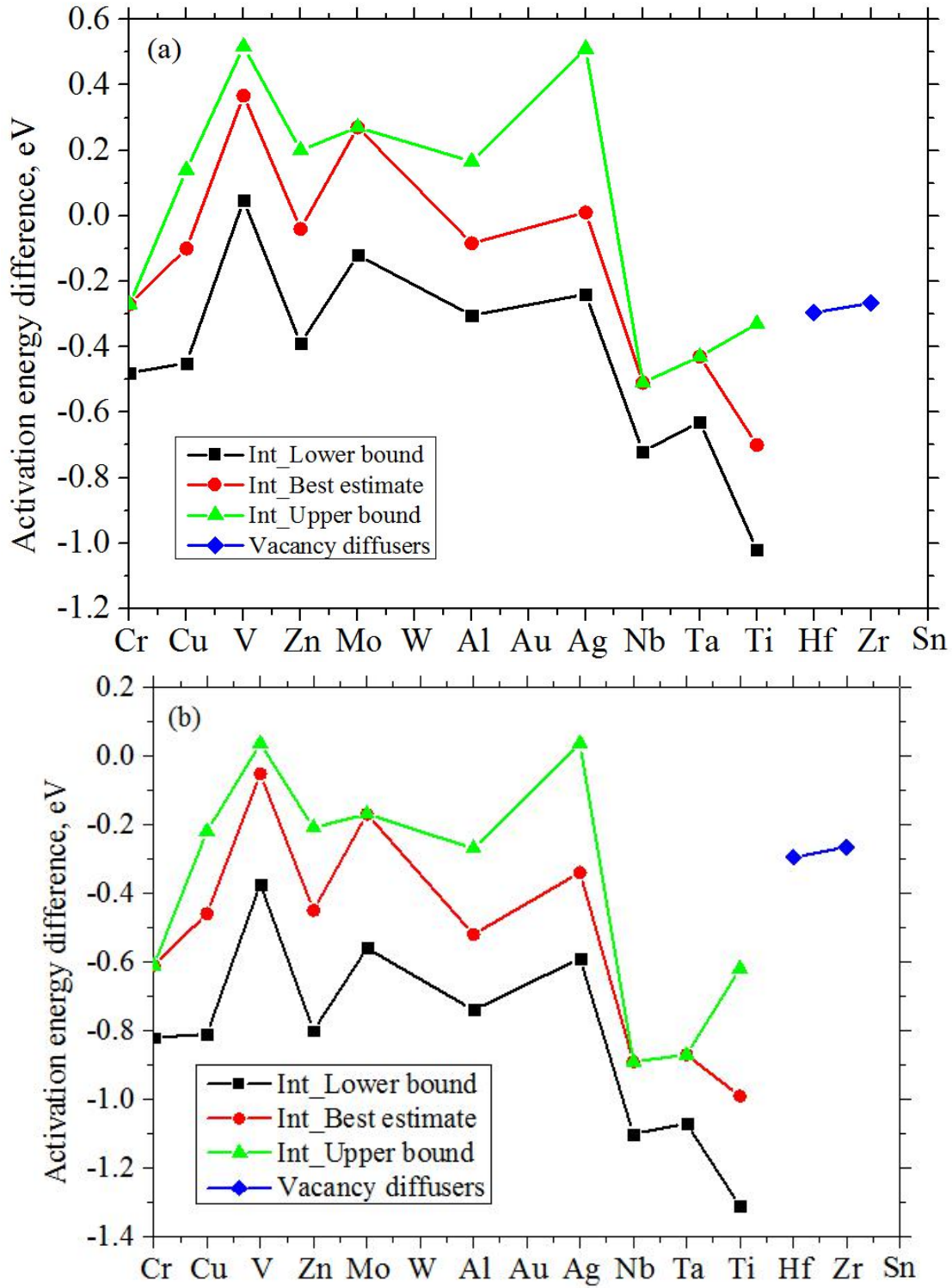

Fig. 6 The difference between the calculated and experimental activation energies at (a) 0 K and (b) 1073 K.

Table 1 The comparison of present and previous DFT calculated vacancy formation energy and lattice parameters of HCP Zr.

| Method | Cell size | $a_0$ (Å) | $c_0$ (Å) | $E_F^V$ (eV) | Ref. |
|---|---|---|---|---|---|
| LDA, SIESTA | 36 | | | 1.98 | [18] |
| LDA, SIESTA | 36 | 3.24 | 5.23 | 2.29 | [45] |
| PBE, SIESTA | 36 | 3.16 | 5.13 | 2.17 | [45] |
| PBE, PWSCF | 200 | | | 2.07 | [46] |
| PW91, VASP | 36 | | | 1.90 | [47] |
| PW91, VASP | 96 | 3.23 | 5.18 | 1.86 | [47] |
| PBE, VASP | 96 | 3.23 | 5.17 | 2.01 | Present work |

Table 2 The calculated solute-vacancy binding energies and migration energies of eight atomic exchanges demonstrated in Fig. 1 for solutes in HCP Zr. The data in bold denote the lowest migration barrier for each solute. Unit in eV.

| Solute | $E_B$ | $E_a$ | $E_b$ | $E_c$ | $E_X$ | $E'_a$ | $E'_b$ | $E'_c$ | $E'_X$ |
|---|---|---|---|---|---|---|---|---|---|
| Cr | -0.23 | 0.54 | 0.36 | 0.25 | 0.66 | **0.10** | 0.52 | 0.91 | 0.77 |
| Cu | 0.06 | 0.43 | 0.45 | 0.70 | 0.65 | **0.32** | 0.40 | 0.59 | 0.83 |
| V | 0.01 | 0.18 | 0.42 | 0.62 | 0.72 | 0.25 | **0.14** | 0.50 | 0.83 |
| Zn | 0.10 | 0.58 | 0.52 | 0.70 | 0.69 | 0.48 | **0.41** | 0.61 | 0.87 |
| Mo | -0.23 | 0.16 | 0.35 | 0.65 | 0.52 | 0.13 | **0.12** | 0.42 | 0.69 |
| W | -0.17 | 0.11 | 0.30 | 0.65 | 0.72 | 0.12 | **0.05** | 0.41 | 0.95 |
| Al | 0.15 | 0.59 | 0.56 | 0.68 | 0.92 | 0.51 | **0.38** | 0.62 | 1.07 |
| Au | 0.03 | 0.51 | 0.48 | 0.71 | 0.70 | **0.39** | 0.40 | 0.61 | 0.92 |
| Ag | 0.06 | 0.63 | 0.54 | 0.73 | 0.65 | 0.51 | **0.46** | 0.63 | 0.86 |
| Nb | -0.07 | 0.32 | 0.44 | 0.62 | 0.55 | 0.37 | **0.27** | 0.54 | 0.69 |
| Ta | -0.04 | 0.32 | 0.43 | 0.62 | 0.78 | 0.37 | **0.28** | 0.54 | 0.98 |
| Ti | 0.03 | 0.47 | 0.54 | 0.59 | 0.80 | 0.48 | **0.42** | 0.58 | 0.85 |
| Hf | 0.04 | 0.63 | 0.56 | 0.55 | 0.75 | 0.62 | **0.53** | 0.57 | 0.86 |
| Sn | 0.05 | 0.70 | 0.60 | 0.75 | 0.79 | 0.62 | **0.41** | 0.69 | 1.01 |

Table 3 The calculated attempt frequencies (Thz) for solute and solvent atoms

| | Cr | Cu | V | Zn | Mo | W | Al | Au | Ag | Nb | Ta | Ti | Hf | Sn |
|---|---|---|---|---|---|---|---|---|---|---|---|---|---|---|
| Solvent | 2.05 | 3.09 | 2.50 | 3.68 | 3.81 | 2.67 | 3.28 | 3.13 | 4.13 | 3.29 | 3.19 | 3.65 | 4.60 | 3.05 |
| Solute | 1.97 | 4.02 | 2.65 | 2.91 | 3.57 | 6.39 | 0.92 | 8.45 | 5.55 | 4.55 | 6.97 | 2.84 | 7.07 | 3.27 |

Table 4 The predicted vacancy-mediated activation energies and pre-exponential factors for 14 solutes in HCP Zr compared with available experimental data.

| Solute | Calculation | | | | Experiment | | $T$ (K) | Ref. |
|---|---|---|---|---|---|---|---|---|
| | $Q_\perp$, $Q_\parallel$ (eV) | | $D_{0\perp}$, $D_{0\parallel}$ ($10^{-5}$ m$^2$/s) | | $Q$ (eV) | $D_0$ ($10^{-5}$ m$^2$/s) | | |
| Cr | 2.45 | 2.55 | 0.83 | 0.84 | 1.31 | 4.9×10$^{-2}$ | 896-1105 | [49] |
| | | | | | ⊥ 1.69 | 2 | 1023-1121 | [50] |
| | | | | | ∥ 1.59 | 2 | 1023-1121 | [50] |
| | | | | | ⊥ 1.62 | 1 | 886-1057 | [51] |
| | | | | | ∥ 1.39 | 0.24 | 886-1057 | [51] |
| Cu | 2.76 | 2.93 | 1.42 | 1.69 | ⊥ 1.60 | 2.5 | 933-1132 | [52] |
| | | | | | ∥ 1.54 | 4 | 888-1132 | [52] |
| | | | | | 1.62 | 4.2 | 887-1117 | [52] |
| V | 2.75 | 2.84 | 1.12 | 1.16 | 0.99 | 1.12×10$^{-7}$ | 873-1123 | [53] |
| Zn | 2.80 | 2.97 | 1.13 | 1.28 | 2.32 | 16.5 | 1002-1114 | [54] |
| | | | | | | | 1099 | [55] |
| Mo | 2.30 | 2.46 | 1.31 | 1.51 | 1.07 | 6.22×10$^{-7}$ | 873-1113 | [56] |
| W | 2.56 | 2.77 | 2.27 | 2.62 | | | | |
| Al | 3.09 | 3.23 | 0.42 | 0.44 | 2.90 | 170 | 873-1073 | [57] |
| | | | | | | | 1108 | [39] |
| Au | 2.73 | 2.95 | 2.37 | 3.18 | | | 1143 | [55] |
| Ag | 2.73 | 2.92 | 1.79 | 2.27 | 1.94 | $5.1\times10^{-2}$ | 1037-1120 | [58] |
| | | | | | ⊥ 1.80 | $5.9\times10^{-3}$ | 1063-1118 | [59] |
| | | | | | ∥ 2.20 | 0.67 | 1063-1118 | [59] |
| | | | | | 2.54 | 22 | 938-1117 | [60] |
| | | | | | 2.18 | 0.68 | 895-1110 | [60] |
| | | | | | 2.11 | 0.17 | 1028-1104 | [61] |
| | | | | | | | 1094 | [55] |
| Nb | 2.52 | 2.66 | 1.60 | 1.80 | 2.69 | 18.8 | 1002-1097 | [62] |
| | | | | | 1.37 | $6.6\times10^{-5}$ | 1013-1130 | [42] |
| Ta | 2.75 | 2.94 | 2.49 | 2.81 | 3.04 | 1000 | 973-1073 | [63] |
| | | | | | | | 872-1096 | [64] |
| Ti | 2.84 | 2.89 | 1.27 | 1.25 | ∥ 2.93 | 170 | 1037-1124 | [65] |
| | | | | | | | 1116 | [39] |
| Hf | 2.80 | 2.90 | 2.72 | 2.83 | 3.10 | 2.6 | 873-1111 | [66] |
| | | | | | | | 773-1115 | [67] |
| | | | | | | | 1095-1096 | [64] |
| | | | | | | | 829-1097 | [41] |
| Sn | 2.84 | 3.06 | 1.19 | 1.40 | 0.95 | $2\times10^{-7}$ | 923-1093 | [68] |

**Table 5 The formation energies (eV) for substitutional and interstitial solutes in HCP Zr at ground state. Some of the values are not available due to the**

instability of the configurations and are marked NA, the data in bold means formation energy for the most stable interstitial site at ground state.

| Solute | $E_F^S$ | $E_H$ | $E_O$ | $E_C$ ($E_{C'}$) | $E_S$ | $E_{BS}$ | $E_{DS}$ | $E_{BO}$ | $E_{BC}$ |
|---|---|---|---|---|---|---|---|---|---|
| Cr | 1.09 | 3.78 | **1.92** | 2.10 | 2.69 | NA | NA | 2.06 | NA |
| Cu | 0.38 | 2.88 | 1.78 | 2.03 | 2.52 | 1.77 | **1.54** | 2.13 | 2.08 |
| V | 0.97 | 3.99 | 2.16 | 2.35 | 2.66 | NA | 2.59 | **2.01** | NA |
| Zn | -0.25 | 3.46 | 1.93 | 2.03 | 2.80 | 1.81 | **1.69** | 2.49 | 2.18 |
| Mo | 1.18 | 3.89 | **2.13** | 2.36 | 2.77 | 2.48 | NA | 2.54 | NA |
| W | 1.39 | 4.69 | **2.75** | 2.99 | 3.39 | 3.14 | NA | 3.21 | NA |
| Al | -0.86 | 4.14 | 2.00 | 1.92 | 2.95 | 1.76 | **1.75** | 2.67 | 2.20 |
| Au | -0.84 | 2.80 | 1.73 | **1.27** | 2.30 | 1.31 | 1.42 | 2.27 | 1.65 |
| Ag | -0.06 | 3.90 | 2.76 | 2.57 | 3.23 | **2.26** | 2.50 | 3.15 | NA |
| Nb | 0.65 | 4.22 | **2.63** | 2.84 | 2.97 | 2.87 | 3.08 | 2.70 | NA |
| Ta | 0.64 | NA | **3.06** | 3.15 | 3.37 | 3.22 | 3.15 | 3.13 | NA |
| Ti | 0.18 | 4.16 | 2.48 | 2.60 | 2.71 | 2.57 | 2.71 | **2.11** | NA |
| Hf | 0.00 | 4.69 | 3.36 | 3.34 | 3.48 | 3.30 | 3.33 | **3.23** | NA |
| Zr | 0.00 | NA | 3.08 | 3.18 | 3.22 | 3.10 | 3.42 | **3.00** | NA |
| Sn | -1.25 | NA | 2.42 | **2.02** | 3.24 | NA | 2.14 | 3.52 | 2.27 |

Table 6 The formation energy difference of the most stable interstitial and substitutional configurations at 0 K and 1073 K. Unit in eV.

| Solute | Interstitial site | 0 K | 1073 K |
|---|---|---|---|
| Cr | O | 0.83 | 0.49 |
| Cu | DS | 1.16 | 0.79 |
| V | BO | 1.03 | 0.62 |
| Zn | DS | 1.94 | 1.52 |
| Mo | O | 0.96 | 0.51 |
| W | O | 1.37 | 0.89 |
| Al | DS | 2.61 | 2.16 |
| Au | C′ | 2.11 | 1.71 |
| Ag | BS | 2.31 | 1.95 |
| Nb | O | 1.98 | 1.59 |
| Ta | O | 2.42 | 1.97 |
| Ti | BO | 1.92 | 1.62 |
| Hf | BO | 3.23 | 2.88 |
| Zr | BO | 3.00 | 2.66 |
| Sn | C | 3.27 | 2.84 |

**Table 7 The interstitial migration energies (eV) for O→O pathways within basal plane and along *c* axis direction**

| | Cr | Cu | V | Zn | Mo | W | Al | Au | Ag | Nb | Ta | Ti | Hf | Zr | Sn |
|---|---|---|---|---|---|---|---|---|---|---|---|---|---|---|---|
| Basal | 0.32 | 0.48 | 0.32 | 0.35 | 0.39 | 0.27 | 0.22 | 0.45 | 0.25 | 0.40 | 0.35 | 0.32 | 0.21 | 0.35 | 0.40 |
| *C* | 0.21 | 0.35 | 0.36 | 0.56 | 0.41 | 0.46 | 0.67 | 0.54 | 0.39 | 0.21 | 0.20 | 0.39 | 0.21 | 0.21 | 1.10 |